\documentclass[12pt]{iopart}
\begin{document}

\title{Lagrangians, gauge functions and Lie groups for 
semigroup of second-order differential equations}

\author{Z. E. Musielak, N. Davachi, M. Rosario-Franco}
\address{Department of Physics, The University of Texas at 
Arlington, Arlington, TX 76019, USA}
\ead{zmusielak@uta.edu}

\begin{abstract}
A set of linear second-order differential equations is converted into a 
semigroup, whose algebraic structure is used to generate novel 
equations.  The Lagrangian formalism based on standard, null and 
nonstandard Lagrangians is established for all members of the semigroup.  
For the null Lagrangians, their corresponding gauge functions are derived.
The obtained Lagrangians are either new or generalization of those previously 
known.  The previously developed Lie group approach to derive some equations
of the semigroup is also described.  It is shown that certain equations of the 
semigroup cannot be factorized and therefore their Lie groups cannot be 
determined.  Possible solution of this problem is proposed and the 
relationship between the Lagrangian formalism and the Lie group 
approach is discussed. 

\end{abstract}


\section{Introduction}

Let $\mathcal{Q}$ be a set of all linear second-order differential equations
(ODEs) of the form $\hat D y(x) = 0$, where $\hat D \equiv {{d^2} / 
{dx^2}} + B(x) {{d} / {dx}} + C(x)$, and $B(x)$ and $C(x)$ are smooth 
($\mathcal{C}^{\infty}$) and ordinary ($B :\ \mathcal{R} \rightarrow 
\mathcal{R}$ and $C :\ \mathcal{R} \rightarrow \mathcal{R}$) functions
to be determined.  In general, $\mathcal{Q}$ contains a wide variety of 
ODEs that can be separated into two subsets, one with known solutions 
and the other with unknown solutions.  According to Murphy [1], there 
are more than five hundreds ODEs in $\mathcal{Q}$ whose solutions 
are known, and many of those solutions are given by the special functions 
(SF) of mathematical physics as defined in [1,2].  
       
An algebraic structure can be added to $\mathcal{Q}$ and as a result the 
set becomes a {\it semigroup}.  Let $\mathcal{S}$ be a semigroup and let
'+' be the binary operation of addition, such that $\forall s_1, s_2 \in 
\mathcal{S} : (s_1 + s_2) \in \mathcal{S}$.  Moreover, the associativity 
axiom requires that $\forall s_1, s_2, s_3 \in \mathcal{S}: s_1 + ( s_2 + 
s_3) = ( s_1 + s_2 ) + s_3$.  With the above conditions being obeyed by 
all elements of $\mathcal{S}$, the semigroup also preserves the commutative 
rule $\forall s_1, s_2 \in \mathcal{S}: (s_1 + s_2 = s_2 + s_1) \in \mathcal{S}$, 
which means that $\mathcal{S}$ is a {\it commutative semigroup}.   It must 
be pointed out that the algebraic structure imposed on the commutative 
semigroup allows generating both new and known ODEs (see Section 2) 
that are also elements of $\mathcal{S}$ . 

Let us also point out that an identity element may also be formally introduced 
into the semigroup $\mathcal{S}$ by taking $B(x) = C(x) = 0$ and finding the 
solution $y(x) = a_0 x + b_0$ with $a_0$ and $b_0$ being constants.  Having 
defined this identity element, the semigroup $\mathcal{S}$ becomes a {\it 
monoid} $\mathcal{M}$, and since the original $\mathcal{S}$ is commutative, 
$\mathcal{M}$ is also commutative.  The monoid structure of the considered 
ODEs is an interesting property that may be studied further in future papers, 
however, in this paper only the commutative semigroup of ODEs is investigated.        

The first main objective of this paper is to establish the Lagrangian formalism
for all ODEs of $\mathcal{S}$ by using standard and non-standard Lagrangians.
The standard Lagrangians (SLs) are typically expressed as the difference between 
terms that can be identified as the kinetic and potential energy [3].  On the other 
hand, for non-standard Lagrangians (NSLs), originally introduced by Arnold [4] 
who referred to them as non-natural Lagrangians, identification of kinetic and 
potential energy terms may not be obvious.   Among different applications, 
Alekseev and Arbuzov [5] used the NSLs to formulate the Yang-Mills field theory.  
Extensive discussions of methods to derive NSLs can be found in some previous 
[6-8] and more recent work [9,10].  Other important methods to obtain NSLs 
were developed by El-Nabulsi [10], Nucci [11], Cari\~nena et al. [12], and 
Saha \& Talukdar [13].  

Lagrangians are not unique, which means that some extra terms may be 
added to them and they give the same original equation.  These are null 
(NLs) or trivial Lagrangians (TLs) as they make the Euler-Lagrange (E-L) 
equation to vanish identically [14,15].  In other words, the NLs can be 
added to any Lagrangian without changing the derivation of the original 
equation.   It is also required that the NLs can be expressed as the total 
derivative of a scalar function, which is called a gauge function [14,16,17].  
The null Lagrangians have been extensively studied in mathematics [18-23]
but only with a couple of applications in physics [24,25].  
  
In theoretical physics, the Lagrangian formalism is central to any classical or 
quantum theory of particles, waves or fields, and the fundamental equations 
of modern physics are typically derived from given Lagrangians [3].  If the 
original equations are given but their Lagrangians are not known, then the 
inverse (or Hemlholtz) problem of the calculus of variations must be solved 
[26].   In this paper, the inverse problem for all ODEs of $\mathcal{S}$ is 
solved and their SLs, NLs and NSLs are derived, the existence of the obtained 
Lagrangians is verified by using the Helmholtz conditions [27], and it is 
demonstrated that the derived NSLs violate the Helmholtz conditions.  

As the second main objective of this paper, the Lie group approach is applied
to the ODEs of $\mathcal{S}$.  A Lie group for a given ODE of $\mathcal{S}$ 
must be known in advance, so its irreducible representations (irreps) can be 
determined and used to obtain the corresponding solutions of this ODE. Moreover, 
operators of the Lie algebra of the selected Lie group can be used to obtain the 
original ODE.  This work has been done so far for the ODEs whose solutions 
are the SF of mathematical physics [28-32] (called here the SF ODEs), and its 
limitations are explored in this paper.  Now, if the original equation is given 
but its Lie group is not known, then the original equation must be factorized 
[33,34] and the resulting lower order ODE must be related to the Lie algebra, 
from which the corresponding Lie group is determined.  In this paper, we 
establish the validity of the Lie group approach to ODEs of $\mathcal{S}$ 
(beyond the SF ODEs) and demonstrate the existence of limits on the 
factorization for some ODEs of $\mathcal{S}$.  

Finally, the Lagrangian formalism and the Lie group approach, the two 
independent methods of obtaining the ODEs of $\mathcal{S}$, are compared.  
The advantages and disadvantages of each method are discussed, and it is 
shown that the Lagrangian formalism can be established for all considered 
ODEs, however, the Lie group approach is only limited to some ODEs that 
form a sub-semigroup of $\mathcal{S}$.  Applications of the obtained 
results to selected problems in mathematical physics are also discussed.

The outline of the paper is as follows: a procedure of generating new 
ODEs in $\mathcal{S}$ is described and applied to some ODEs in 
Section 2; the Lagrangian formulation for all ODEs of $\mathcal{S}$ 
is established by using standard and non-standard Lagrangians and 
applied to selected ODEs in Section 3; the Lie group approach and 
factorization methods are presented in Section 4; comparison between 
the two methods is discussed in Section 5; and conclusions are given 
in Section 6.    

\section{Commutative semigroup of differential equations}   

An interesting property of $\mathcal{S}$ is that its algebraic operation can be used 
to generate novel ODEs by simply adding two different elements of the semigroup.  
Since any two elements (two chosen ODEs) of $\mathcal{S}$ can be added up, the 
results may be rather surprising as shown by a couple of examples below.  Let 
$D_1 = {{d^2} / {dx^2}} + B_1 (x) {{d} / {dx}} + C_1 (x)$ and $\hat D_2 = 
{{d^2} / {dx^2}} + B_2 (x) {{d} / {dx}} + C_2 (x)$, where $B_1 (x)$, $B_2 (x)$, 
$C_1 (x)$ and $C_2 (x)$ are smooth and arbitrary functions, be two operators, and 
let $\hat D_1 y(x) = 0$ and $\hat D_2 y(x) = 0$ be two ODEs of $\mathcal{S}$.  
Then, the binary operation of $\mathcal{S}$ allows writing
\begin{equation}
\hat D y (x) = (\hat D_1 + \hat D_2) y(x) = 0\ ,
\label{S2eq1}
\end{equation}
where $B(x) = [B_1(x) + B_2(x)] / 2$ and $C(x) = [C_1(x) + C_2(x)] / 2$.

Here are some interesting examples.  All Bessel (regular, modified, 
spherical and modified spherical) equations are elements of $\mathcal{S}$.
To add the regular and modified Bessel equations, we have $B_1(x) = 
B_2(x) = 1 / x$, $C_1(x) = 1 - \mu^2 / x^2$ and $C_2(x) = - (1 + \mu^2 
/ x^2)$, where $\mu$ is either real or integer constant, which gives 
\begin{equation}
y^{\prime \prime} (x) + {{1} \over {x}} y^{\prime} (x) - {{\mu^2} \over 
{x^2}} y (x) = 0\ .
\label{S2eq2a}
\end{equation}
Since the eigenvalue of this equation is $\lambda = 0$ (see Section 4), the 
resulting ODE is neither regular nor modified Bessel equation but instead 
it is known as the Euler equation [1].   Addition of other Bessel equations 
gives novel Bessel and Euler equations.

To show these novel equations, let a general form of Bessel equation be  
\begin{equation}
y^{\prime \prime} (x) + {{\alpha} \over {x}} y^{\prime} (x) + 
\left ( \beta - {{\mu^2} \over {x^2}} \right ) y (x)
= 0\ ,
\label{S2eq2b}
\end{equation}
where $B(x) = \alpha / x$ and $C(x) = \beta (1 + \gamma \mu^2 
/ x^2$).  In addition, $\alpha$ and $\beta$ are constants, whose 
values correspond to four Bessel equations and the Euler equation 
(see Table 1).  In addition, $\mu$ is either integer or real.  Note 
that the form of Eq. (\ref{S2eq2}) considered here is more general 
than that used in [10] as it allows reducing the general Bessel equation
to the Euler equation [1]. 
\begin{table}
\center{
\caption{Bessel and Euler equations}
\footnotesize
\begin{tabular}{@{}llll}
\br
\textbf{Known and derived equations} & \boldmath{$\alpha$} & \boldmath{$\beta$}\\
\mr
 Regular Bessel                                 & 1 & 1\\
 Modified Bessel                                & 1 & -1\\
 Spherical Bessel                               & 2 & 1\\
 Modified spherical Bessel                   & 2 & -1\\
 Semi-spherical Bessel$^*$                & 3/2 & 1\\
 Modified semi-spherical Bessel$^*$   & 3/2 & -1\\
 Regular Euler                                   & 1 & 0\\
 Spherical Euler$^*$                         & 2 & 0\\
 Semi-spherical Euler$^*$                 & 3/2 & 0\\
\br
\end{tabular}\\
$^{*}$Novel equations derived by using the binary operation of $\mathcal{S}$.}
\end{table}
\normalsize

Note that for $\alpha = 3/2$, a new name 'semi-spherical' was introduced 
for both Bessel and Euler equations.  Additions of Bessel equations required to 
obtain the Euler equation and the novel equations (see Table 1) are presented
in Table 2.  It must be pointed out that further additions of the Bessel equations 
and the derived equations are also possible but not considered here. 

\begin{table}
\center{
\caption{Derivation of novel Bessel and Euler equations}
\footnotesize
\begin{tabular}{@{}llll}
\br
\textbf{Derived equations} & \textbf{Addition of Bessel equations}\\
\mr
 Semi-spherical Bessel                   & Regular and spherical\\
 Modified semi-spherical Bessel      & Modified and modified spherical\\
 Regular Euler                              & Regular and modified\\
 Spherical Euler                            & Spherical and modified spherical\\
 Semi-spherical Euler                    & Regular and modified spherical\\
 Semi-spherical Euler                    & Modified and spherical\\
\br
\end{tabular}\\
$^{*}$Novel equations derived by using the binary operation of $\mathcal{S}$.}
\end{table}
\normalsize

Another interesting result is obtained when the regular Legendre equation
with $B_1(x) = - 2x / (1 - x^2)$ and $C_1(x) = l (l + 1) / (1 - x^2)$ is 
added up together with the associated Legendre equation with $B_2(x) 
= - 2x /(1 - x^2)$ and $C_1(x) = l (l + 1) / (1 - x^2) - m^2 / (1 - x^2)^2$,
where $l$ and $m$ are constants.  The resulting ODE is again the associated 
Legendre equation given by 
\begin{equation}
y^{\prime \prime} (x) - {{2x} \over {(1-x^2)}} y^{\prime} (x) + 
\left [ {{l (l+1)} \over {(1-x^2)}} - {{\bar m^2} \over {(1-x^2)^2}}
\right ] y (x) = 0\ ,
\label{S2eq3}
\end{equation}
where $\bar m^2 = m^2 / 2$.  This shows that only the constant 
$\bar m$ is affected by the addition but the form of the resulting ODE
does not change.

The considered simple examples show that numerous novel ODEs 
can be generated by using the algebraic operation of the semigroup 
$\mathcal{S}$.  Since the process of adding the elements of 
$\mathcal{S}$ to each other can be carried on {\it ad infinitum}, 
there are infinite different resulting ODEs, and all of them are the 
elements of $\mathcal{S}$.

\section{Lagrangian formalism and its Lagrangians}

Let $\mathcal{J} [y(x)]$ be a functional that depends on an ordinary 
and smooth function $y(x)$, and let $\mathcal{J} [y(x)]$ be defined 
by an integral over a smooth function $L$, which is called Lagrangian,
and depends on $y'(x) =  dy / dx$, $y$ and on $x$, so $L (y', y, x)$.
The Principle of Least Action or Hamilton's Principle [3] require that 
$\delta \mathcal{J} = 0$, where $\delta$ is the variation defined as 
the Fr\'echet derivative of $\mathcal{J} [y(x)]$ with respect to $y(x)$.  
Using the condition $\delta \mathcal{J} = 0$, the E-L equations are 
obtained and their solutions give $y(x)$ that makes the action stationary.  
The described procedure is the basis of the classical calculus of variations 
and it works well when the Lagrangian $L (y', y, x)$ is either already 
given or it can be determined for a given physical system.  

The basis for the Lagrange formalism is the jet-bundle theory [20,21].
Let $X$ and $Y$ be differentiable manifolds of dimensions $m$ and 
$M+m$, respectively, and let $\pi: Y \rightarrow X$ be a fibred bundle 
structure, with $\pi$ being the canonical projection of the fibration.  
Let $J^r_m (Y) \rightarrow X$ be the r-th jet bundle, with $x \in X$,
$y \in Y$ and $r \in IN$.  Then, an ODE of order $q$ is called locally 
variational (or the E-L type) if, and only if, there exists a local real 
function $L$ constrainted by the condition $q \leq r$.  For the ODEs 
of $\mathcal{S}$, $q = 2$ and $L (y', y, x)$ is a local Lagrangian. 
Such local Lagrangians are not unique as other Lagrangians may 
also exist and they would give the same original equations when 
substituted into the E-L equations.  Now, if Lagrangians are not 
known, then the problem of finding them is the inverse (or Helmholtz) 
variational problem [26,27,35].  One of the main goals of this paper 
is to solve this problem for all ODEs of $\mathcal{S}$.

\subsection{Standard Lagrangians}

In the previous work [6,7], special forms of standard Lagrangians 
$L_{s} [y'(x), y(x), x]$ were derived for $\hat D y(x) = 0$.  These
results aer generalized in Proposition 1 by allowing the Lagrangians 
to depend on smooth and ordinary functions that are to be determined.\\

\noindent
{\bf Proposition 1.}  {\it Let $L_{s} [y'(x), y(x), x]$ be the standard Lagrangian 
given by 
\begin{equation}
L_{s} [y'(x), y(x), x] = {1 \over 2} \left [ f_1 (x) \left ( y' (x) \right )^2 +
f_2 (x) y' (x) y (x) + f_3 (x) y^2 (x) \right ]\ ,
\label{S3eq1}
\end{equation}
and let $f_1(x)$, $f_2(x)$ and $f_3(x)$ be smooth and ordinary functions 
to be determined.  The Lagrangian $L_{s} [y'(x), y(x), x]$ can be used to
obtain the original equation $\hat D y (x) = 0$, if and only if, $f_1 (x) > 0$, 
$f_{1}^{\prime} (x) = B (x) f_1 (x)$ and $f_2^{\prime} (x) / 2 - f_3 (x) = 
C (x) f_1 (x)$.}\\

\noindent
{\bf Proof.}  Substitution of $L_{s} [y'(x) , y(x), x]$ into the E-L 
equation gives
\begin{equation}
f_1 (x) y^{\prime \prime} (x) + f_1^{\prime} (x) y' (x) + [ f_2^{\prime} 
(x) / 2 - f_3 (x) ] y (x) = 0\ ,
\label{S3eq2}
\end{equation}
or 
\begin{equation}
y^{\prime \prime} + {{f_1^{\prime}} \over {f_1 (x)}} y' (x) + 
{1 \over {f_1 (x)}} \left [ f_2^{\prime} (x) / 2 - f_3 (x) \right ] y (x) 
= 0\ ,
\label{S3eq3}
\end{equation}
if $f_1 (x) > 0$.  This E-L equation can be converted into the original
equation $y^{\prime \prime} + B (x) y' (x) + C (x) y (x) = 0$ if  
$f_{1}^{\prime} (x) = B (x) f_1 (x)$ and  $f_2^{\prime} (x) / 2 -
f_3 (x) = C (x) f_1 (x)$, which concludes the proof.\\

Other implications of the results of Proposition 1 are now presented 
by the following corollaries.\\  

\noindent
{\bf Corollary 1.} {\it The function $f_1 (x)$ is given as $f_1 (x) = 
a_1 E_s (x)$, where $a_1$ is an integration constant and $E_s (x) 
= \exp{ [ \int^x B(\tilde x) d \tilde x ]}$.}\\ 

\noindent
{\bf Corollary 2.} {\it The Euler-Lagrange equations do not supply enough
constraints to determine both $f_2 (x)$ and $f_3 (x)$, thus, one of these 
functions must be specified.  Different choices of $f_2 (x)$ or $f_3 (x)$ 
lead to different standard Lagrangians.}\\

For a given physical problem with specified intitial conditions, more 
constraints on the functions $f_2 (x)$ and $f_3 (x)$ can be imposed.
However, in this paper, we keep our analysis general and do not
consider any specific physical problem with given initial conditions.

We may use the result of Corollary 2 to define three Lagrangians that 
play special roles in this paper.  In general, many different Lagrangians 
can be obtained but considering all of them is out of the scope of this 
paper.  Let us begin with the choice of $f_2 (x) = 0$, which sets $f_3 (x) 
= - C (x)$.  The result is a Lagrangian that is called here the minimal 
Lagrangian $L_{s,min} [y'(x), y(x), x]$ and it is defined as
\begin{equation}
L_{s,min} [y'(x), y(x), x] = {a_1 \over 2} \left [ \left ( y' (x) \right )^2 
- C (x) y^2 (x) \right ] E_s (x)\ .
\label{S3eq4}
\end{equation}
It is easy to verify that this Lagrangian is the simplest standard Lagrangian 
that allows obtaining the original equation $\hat D y (x) = 0$ directly from 
the E-L equations [6].  Let us also point out that the same Lagrangian is 
obtained when $f_2 (x) y' (x) + f^{\prime}_2 (x) y (x) / 2 = 0$, which 
requires that $f_2 (x)$ depends on $y(x)$ and that their relationship 
is $f_2 (x) = [ \bar a_1 / y^2 (x) ]^2$, where $\bar a_1$ is another 
integration constant. 

We may also choose $f_2 (x) = a_2$ = const, which gives $f_3 (x) = 
- C (x)$ and the so-called middle Lagrangian $L_{s,mid} [y'(x), y(x), x]$ 
is obtained
\begin{equation}
L_{s,mid} [y'(x), y(x), x] = L_{s,min} [y'(x), y(x), x] + L_{o,mid} 
[y'(x), y(x), x]\ ,
\label{S3eq5}
\end{equation}
where $L_{o,mid} [y'(x), y(x), x]= (a_2 / 2) y' (x) y (x)$ is a null 
Lagrangian [10].  The middle Lagrangian is a new standard Lagrangian.

Finally, we take $f_2 (x) = f^{\prime}_1 (x) = a_1 B (x) E_s (x)$, which gives 
$f_3 (x) = f_2^{\prime} (x) / 2 - C (x) f_1 (x) = ( a_1 / 2 ) [ B' (x) + B^2 (x) 
- 2 C (x) ] E_s (x)$, and the maximal Lagrangian $L_{s,max} [y'(x), y(x), x]$
can be written as  
\begin{equation}
L_{s,max} [y'(x), y(x), x] = L_{s,min} [y'(x), y(x), x] + L_{o,max} [y'(x), y(x), x]\ ,
\label{S3eq6}
\end{equation}
where $L_{o,max} [y'(x), y(x), x] = (a_1 / 2) [ B (x) y' (x) + \left ( B' (x) + B^2(x) 
\right ) y(x) / 2 ] y(x) E_s (x)$ is another null Lagrangian.  The Lagrangian $L_{s,max} 
[y'(x), y(x), x]$ was also previously found [7], and it was demonstrated that this 
Lagrangian gives the original equation upon substitution into the E-L equation.  
It is easy to verify that even if the obtained Lagrangians are different they lead
to the same original equation $\hat D y (x) = 0$.

\subsection{Null Lagrangians}

A well-known result is that the difference between two Lagrangians that give
the same equation can be written as the total derivative of scalar function 
$\Phi (y,x)$ [15].  This means that we may write 
\begin{equation*}
L_{s,mid} [y'(x), y(x), x] - L_{s,min} [y'(x), y(x), x] = L_{o,mid} 
[y'(x), y(x), x] = {{d \Phi_{1}} \over {dx}}
\end{equation*}
\begin{equation*}
L_{s,max} [y'(x), y(x), x] - L_{s,min} [y'(x), y(x), x] = L_{o,max} 
[y'(x), y(x), x] = {{d \Phi_{2}} \over {dx}}
\end{equation*}
and
\begin{equation*}
L_{o,max} [y'(x), y(x), x] - L_{o,mid} [y'(x), y(x), x] = 
{{d \Phi_{3}} \over {dx}}\ ,
\end{equation*}
where the functions $\Phi_1 (y,x)$, $\Phi_2 (y,x)$ and $\Phi_3 (y,x)$ 
exist on the configuration manifold $M$ (see Section 2.1), however, it 
may not always be possible to find a single-value function on the entire 
$M$ [15].  Since the null Lagrangians can be added to any Lagrangian
without modifying the resulting original equations, we may call them 
'gauge transformations' in reference to the gauge transformation of
electromagnetism.  Moreover, the above results demonstrate that 
the functions $\Phi_1 (y,x)$, $\Phi_2 (y,x)$ and $\Phi_3 (y,x)$ are 
the gauge functions, and that they exist [15].\\

It is easy to find $\Phi_1 (y) = a_2 y^2 / 4 + a_3$, 
where $a_3$ is an integration constant, however, the other 
two functions require $B(x)$ to be specified, which will not 
be done here to keep our approach as general as possible.
Important consequences of the fact that the null Lagrangians 
allow finding the gauge functions is now summarized in the 
following corollaries.\\ 

\noindent
{\bf Corollary 3.} {\it The null Lagrangians $L_{o,mid} [y'(x), 
y(x), x]$ and $L_{o,max} [y'(x), y(x), x]$ are the solutions that 
satisfy the E-L equations but cannot be used to derive $\hat D 
y (x) = 0.$}\\

\noindent
{\bf Corollary 4.} {\it The null Lagrangians $L_{o,mid} [y'(x), y(x), x]$ 
and $L_{o,max} [y'(x), y(x), x]$ can be added to any known Lagrangian 
without making any changes in the resulting original equation.}\\

Finally, let us point out that both $L_{o,mid} [y'(x), y(x), x]$ and $L_{o,max} 
[y'(x), y(x), x]$ depend on $B(x)$ and $y(x)$ but are independent of $C(x)$.
Since the first-derivative term with its $B(x)$ coefficient can be removed 
from $\hat D y(x) = 0$ by using the integral transformation given by 
Proposition 5 (see Section 3.2), the null Lagrangians for such transformed 
ODEs would be zero.  Our results show that with $L_{o,mid} [y'(x), y(x), x] 
= 0$ and $L_{o,mid} [y'(x), y(x), x] = 0$, the {\it minimal Lagrangian} is the 
{\it simplest and most fundamental standard Lagrangian} for all ODEs of 
$\mathcal{S}$.  

\subsection{Non-standard Lagrangians}

We consider non-standard Lagrangians to be given in the following general
form [9,10]
\begin{equation}
L_{ns} [y'(x) , y(x), x] = {1 \over {g_1 (x) y^{\prime} (x) + g_2 (x) 
y (x) + g_3 (x)}}\, 
\label{S3eq13}
\end{equation}
where $g_1 (x)$, $g_2 (x)$ and $g_3 (x)$ are smooth functions to
be determined.  Comparing $L_{ns} [y'(x) , y(x), x]$ to $L_{s} [y'(x), 
y(x), x]$ given by Eq. (\ref{S3eq1}), it is seen that both Lagrangians
have three arbitrary functions, however, the forms of these Lagrangians
are significantly different.  

Studies of the Lagrangian $L_{ns} [y'(x) , y(x), x]$ previously performed 
[8] were limited to only a few special cases that allowed uniquely determine 
the functions $g_1 (x)$, $g_2 (x)$ and $g_3 (x)$.  These functions were 
determined by substituting $L_{ns} [y'(x) , y(x), x]$ into the E-L equations.  
First, it was shown that $g_3 (x)$ separates from the other two functions 
and, as a result, it can be evaluated independently.  Second, the evaluation 
of $g_1 (x)$ and $g_2 (x)$ is connected and they can only be obtained if 
a Riccati equation is solved but neither [7] nor [8] gives the required solutions.  
Therefore, here we solve the resulting Riccati equation [9] in the following 
proposition.\\

{\bf Proposition 2.}  {\it Let  
\begin{equation}
u^{\prime}   + {1 \over 3} u  ^2 - {1 \over 3} u   B (x) - \left [ 
{2 \over 3} B^2 (x) + 2 B'(x) - 3 C (x) \right ] = 0\ , 
\label{S3eq14}
\end{equation}
be the Riccati equation, and let $u(x) = g^{\prime}_1 (x) / g_1 (x)$, 
where $g_1 (x)$ is an arbitrary function of the nonstandard Lagrangian
given by Eq. (\ref{S3eq13}).  The solution of this Riccati equation is
\begin{equation}
u (x) = 3 {{\bar v^{\prime} (x)} \over {\bar v (x)}} + 2 B(x)\ ,
\label{S3eq15}
\end{equation}
if, and only if, the function $\bar v(x)$ satisfies $\hat D \bar v(x) = 0.$}\\ 

\noindent
{\bf Proof.}  We transform the Riccati equation by introducing a new variable 
$v (x)$, which is related to $u (x)$ by $u (x) = 3 v^{\prime} (x) / v (x)$ with 
$v(x) \neq 0$, and obtain
\begin{equation}
v^{\prime \prime} + B(x) v^{\prime} + C(x) v = F (v^{\prime}, 
v, x)\ ,
\label{S3eq16}
\end{equation}
where $F (v^{\prime}, v, x) = 2\ [ 2 B(x) v^{\prime} + B'(x) 
v + B^2 (x) v / 3 ] / 3$.  

Now, we transform Eq. (\ref{S3eq16}) by using 
\begin{equation}
v (x) = \bar v (x)\ \exp{\left [ \int^x \chi(\tilde x) d \tilde x 
\right ]}\ ,
\label{S3eq17}
\end{equation}
which gives $\chi(x) = - 2 B(x) / 3$ [10], and the solution is
\begin{equation}
u (x) = 3 {{\bar v^{\prime} (x)} \over {\bar v (x)}} + 2 B(x)\ ,
\label{S3eq18}
\end{equation}
\noindent
where $\bar v(x)$ must satisfy $\hat D \bar v = 0$.  This concludes
the proof.\\

Having obtained the solution of the Riccati equation, we may use it
to calculate the functions $g_1 (x)$ and $g_2 (x)$ and find the 
following non-standard Lagrangian 
\begin{equation}
L_{ns} [y^{\prime} (x), y(x), x] = {{E_{ns} (x)} \over {[y^{\prime} (x) 
\bar v (x) - y (x) \bar v^{\prime} (x)]\ \bar v^2 (x)}}\ ,   
\label{S3eq19}
\end{equation}
where $E_{ns} (x) = \exp{ [ - 2 \int^x B(\tilde x) d \tilde x ]}$.  It is 
seen that the derived Lagrangian depends explicitly on $\bar v (x)$,
which is a solution to $\hat D \bar v(x) = 0$ that becomes an auxiliary 
condition for the developed Lagrangian formalism based on non-standard
Lagrangians.  

The obtained non-standard Lagrangians were previously derived but 
only for some ODEs whose solutions are special functions of mathematical 
physics [6,8,10].  Those previous results are here generalized to all ODEs 
of $\mathcal{S}$.  Both standard and non-standard Lagrangians allow 
obtaining the original equations, however, the non-standard Lagrangians 
lead to a new phenomenon in calculus of variations, which is the requirement 
of the auxiliary condition in order to obtain $\hat D y(x) = 0.$  

\subsection{Applications to Bessel and Euler equations}

We now derive Lagrangians for the general Bessel equation given by 
Eq. (\ref{S2eq2b}).  More specifically, we find the minimal, middle and 
maximal standard Lagrangians using Eqs. (\ref{S3eq4}), (\ref{S3eq5}) 
and (\ref{S3eq6}), respectively, as well as the non-standard Lagrangian 
given by Eq. (\ref{S3eq19}).   

The minimal (basic) standard Lagrangian is given by 

\begin{equation}
L_{s,min} [y'(x), y(x), x] = {1 \over 2} \left [ \left ( y' (x) \right )^2 
- \left ( \beta - {{\mu^2} \over {x^2}} \right ) y^2 (x) \right ] 
x^{\alpha}\ ,
\label{S3eq20}
\end{equation}
where $a_1 = 1$ in Eq. (\ref{S2eq4}).

The middle Lagrangian can be written as 
\begin{equation}
L_{s,mid} [y'(x), y(x), x] = L_{s,min} [y'(x), y(x), x] + L_{o,mid} 
[y'(x), y(x), x]\ ,
\label{S3eq21}
\end{equation}
where
\begin{equation}
L_{o,mid} [y'(x), y(x), x] = {1 \over 2} y' (x) y (x)\ ,
\label{S3eq21}
\end{equation}
with $a_2 = 1$ and $L_{o,mid} [y'(x), y(x), x]$ being 
a null Lagrangian that does not depend on either $\alpha$
or $\beta$.

The maximal Lagrangian is
\begin{equation}
L_{s,max} [y'(x), y(x), x] = L_{s,min} [y'(x), y(x), x] + 
L_{o,max} [y'(x), y(x), x]\ ,
\label{S3eq22}
\end{equation}
where 
\begin{equation}
L_{o,max} [y'(x), y(x), x] = {{\alpha} \over {2x}} \left [ y' (x) 
+ {{1} \over {2 x}} (\alpha - 1) y(x) \right ] x^{\alpha} y(x)\ ,
\label{S3eq23}
\end{equation}
is another null Lagrangian that depends on $\alpha$ but is independent 
of $\beta$, which is only present in the basic Lagrangian.

According to Eq. (\ref{S3eq19}), the non-standard Lagrangian becomes 
\begin{equation}
L_{ns} [y^{\prime} (x), y(x), x] = {{x^{-2 \alpha}} \over {[y^{\prime} (x) 
\bar v (x) - y (x) \bar v^{\prime} (x)]\ \bar v^2 (x)}}\ ,   
\label{S3eq24}
\end{equation}
with the auxiliary condition $\hat D \bar v (x) = 0$.  It is seen that 
$L_{ns} [y^{\prime} (x), y(x), x]$ depends on $\alpha$ but does  
not dependent explicitly on $\beta$; the $\beta$-dependence is 
only through the solution for $\bar v (x)$.  

The obtained Lagrangians are expressed in terms of two parameters 
$\alpha$ and $\beta$ that uniquely determine whether the considered 
equation is the Bessel or Euler equation (see Table 1).  In other words, 
the standard and non-standard Lagrangians for each equation can be
obtained from Eqs. (\ref{S3eq20}) through (\ref{S3eq24}) by using 
the values of $\alpha$ and $\beta$ given in Table 1.

\subsection{Helmholtz conditions}

Our results show that the Lagrangian formalism may be established for
all ODEs of $\mathcal{S}$ by using either the standard or non-standard
Lagrangians.  The existence of the obtained Lagrangians must be validated
by the Helmholtz conditions [27].  There are three original Helmholtz 
conditions and two of them are trivially satisfied.  The third condition
requires a lot of caution because its original version shows that no 
Lagrangian can be constructed for the general form of the ODEs used
in this paper, namely, $\hat D y(x) = 0$.  It is easy to demonstrate 
that $\hat D y(x) = 0$ is inconsistent with the third Helmholtz condition 
[14].  

To fix the problem of validity of the third Helmholtz condition, we 
substitute either the minimal, middle or maximal Lagrangian into
the E-L equation and, as expected, all three Lagrangians give the 
same original equation.  The resulting ODE is $[\hat D y(x)] E_s 
(x) = 0$ and this equation is {\it consistent} with the third Helmholtz 
condition.  The problem arises when one more step is performed,
namely, dividing the equation by $E_s (x)$, which gives the original
ODE but the resulting ODE violates the third Helmholtz condition;
note that in principle the division can be performed because in 
general $E_s (x) \neq 0$.  In other words, the third Helmholtz 
condition tells us that Lagrangian exists for $[\hat D y(x)] E_s 
(x) = 0$ but it does not exist for $\hat D y(x) = 0$, however, 
it is the latter that we want to derive as it is our original equation.
Thus, the origin of the inconsistency is clearly demonstrated.

The problem with the non-standard Lagrangians is more profound
because they do allow obtaining the original equation but the 
resulting ODE does not obey the third Helmholtz condition.  There 
are a few special cases that may lead to a result that is consistent 
with this condition.  However, for the original equation $\hat D y(x) 
= 0$, which is derived from the non-standard Lagrangians, the 
results remain inconsistent with the third Helmholtz condition.  This 
violation of the third Helmholtz condition is a known phenomenon
in the calculus of variations [10].

\section{Lie group approach}

\subsection{Special function equations}

Linear second-order ODEs whose solutions are special functions (SFs) 
of mathematical physics are called here the special function equations 
(SFEs).  All SFEs are elements of $\mathcal{S}$ and they form a 
sub-semigroup denoted $\mathcal{S}_{sf}$, with $\mathcal{S}_{sf} 
\in \mathcal{S}$.  There are many applications of SFEs and SFs in 
applied mathematics, physics and engineering [2,29].  The Lie group 
approach provides a unifying perspective of all SFs and their SFEs.
Therefore, it is surprising that the approach is not commonly known 
and that its descriptions can only be found in a very few advanced 
texbooks of mathematical methods (e.g., [29]) or monographs [30,
31] and reviews [32] devoted to this topic. 

Using the previously obtained results [16-19], we find that Bessel and 
spherical Bessel functions are obtained respectively from the Euclidean 
groups $E(2)$ and $E(3)$, also known as the ISO(2) and ISO(3) groups.  
In addition, it was shown that Legendre polynomials and functions are 
derived respectively from unitary groups $SU(2)$ and unimodular 
quasi-unitary groups $QU(2)$.  As long as Hermite polynomials and 
functions are concerned, they can be obtained from Heisenberg 
$H_3$ and $H_4$ groups, respectively.
 
Let us now demonstrate how the procedure of finding special functions 
works by selecting the $E(2)$ group, which consists a subgroup of all 
translations $T(2)$ in $\mathcal{R}^2$, and a subgroup of all rotations 
$R(2)$ in $\mathcal{R}^2$;  thus, the group structure is $E(2) = R(2) 
\otimes_s T(2)$, where $T(2)$ is an invariant subgroup.  The procedure
requires finding a function that transform as one of irreps of a given group;
in the previous work [36], such function was found for the Galilean group 
of the metric, however, here it is obtained for the Euclidean group $E(2)$
in the following proposition and corollary.\\

\noindent
{\bf Proposition 3.}  {\it Let $\hat T_{\vec a}$ be a translation operator 
of $T(2) \in E(2)$ and let $\phi (\vec r)$ be a smooth and ordinary function.  
The function $\phi (\vec r)$ transforms as the irreps of $T(2)$ if, and only 
if, $\phi (\vec r)$ satisfies 
the eigenvalue equations 
\begin{equation}
i \nabla\phi (\vec r) = \vec k \phi (\vec r)\ ,
\label{S4eq1}
\end{equation}
where $\vec k$ labels the irreps.}\\

\noindent
{\bf Proof}.  The action of $\hat T_{\vec a}$ on $\phi (\vec r)$ 
is given by
\begin{equation}
\hat T_{\vec a} \phi (\vec r) = \phi (\vec r + \vec a) = 
\exp [ i \vec k \cdot \vec a ] \phi (\vec r)\ ,
\label{S4eq2}
\end{equation}
where $\vec a$ is a translation and $\vec r = (x,y)$.  Making 
the Taylor series expansion of $\phi (\vec r + \vec a)$, we 
find
\begin{equation}
\phi (\vec r + \vec a) = \exp [ i (- i \vec a \cdot \nabla) ] 
\phi (\vec r)\ .
\label{S4eq3}
\end{equation}
Comparing Eq. (\ref{S4eq3}) to Eq. (\ref{S4eq2}), we 
obtain the following eigenvalue equations 
\begin{equation}
i \nabla\phi (\vec r) = \vec k \phi (\vec r)\ ,
\label{S4eq4}
\end{equation}
which represent the necessary conditions that $\phi (\vec r)$
transforms as the irreps of $T(2)$ [35].  This concludes the proof.\\   

\noindent
{\bf Corollary 5.} {\it The solutions to the eigenvalue equations of 
Proposition 4 are the plane waves $ \phi (\vec r) = \phi_0 e^{i \vec 
k \cdot \vec r}$, where $\phi_0$ is an integration constant.}\\

By performing the Fourier-sine and Fourier-cosine expansions of the 
plane waves of Corollary 5, one finds that each expansion has only 
two non-zero coefficients, and that one of these coefficients is 
proportional to $J_{\mu} (x)$, and the other to $J_{-\mu} (x)$, with 
$\mu$ being the label of the coefficients and the resulting SFs being 
Bessel functions [2,29].  Moreover, the Lie algebra of the $E(2)$ group 
allows finding the raising $\hat A_{+} (x, \mu) J_{\mu} (x) = [ d/dx - 
(\mu - 1/2)/x] J_{\mu} (x) = J_{\mu + 1} (x)$ and lowering $\hat 
A_{-} (x, \mu) J_{\mu} (x) = [ - d/dx - (\mu - 1/2)/x] J_{\mu} (x) 
= J_{\mu - 1} (x)$ operators.  Using these operators, the power 
series expansions for $J_{\mu} (x)$ and $J_{-\mu} (x)$, their 
recurrence relationships, generating functions, and other properties 
may be obtained, without making any reference to the Bessel equation 
[30 - 32].  To derive the Bessel equation, $\hat A_{+} \hat A_{-} J_{\mu} 
(x) = \hat A_{-} \hat A_{+} J_{\mu} (x) = J_{\mu} (x)$ may be used, 
however, it must be noted that the resulting ODE is the Bessel equation 
written in its second canonical form (see Proposition 4). 

As described above, the remarkable relationships between Lie groups 
and the SFs of the ODEs of $\mathcal{S}_{sf}$ have been found.  The 
relationships are important as they unified the well-known and commonly 
used SFs with simple Lie groups used in physics and applied mathematics.  
To the best of our knowledge, the Lie group approach has only been 
applied to the ODEs of $\mathcal{S}_{sf}$, and Lie groups for these 
equations have been identified.  However, the semigroup $\mathcal{S}$ 
contains also many other ODEs, which we denote here as $\mathcal{S}_{re}$,
with the subscript standing for the remaining ODEs.  In general, the solutions 
of  $\mathcal{S}_{re}$ are either known, but not given in terms of the SFs, 
or unknown.  In the following, we determine whether the Lie group approach 
can also be applied to all ODEs of $\mathcal{S}_{re}$, or not.

\subsection{Remaining equations}

Now, if a Lie group for a given ODE is not known, then one may factorize the 
ODE, obtain the eigenvalue equations and the raising and lowering operators 
[15] that are related to a Lie algebra of infinitesimal group generators from 
which the corresponding Lie group can be determined [30 -32].  Using the 
obtained Lie algebra, the SFs naturally appear as the basis functions of 
differential operator representations of the algebra.  Moreover, once the 
Lie group is known, its irreps can be found, and if $\Gamma$ is the matrix 
of such irreps, then their matrix element $\gamma_{ij} (g)$, where $g$ is a 
group element, becomes proportional to one of the SFs, or to a product of 
SFs.  This shows that the Lie group approach is established only for those 
ODEs that are factorized.  There are many factorization methods (e.g., [33]
and references therein), however, only the Infeld and Hull (IH) factorization 
[34] shows clear relationships to Lie algebra [30 - 32], therefore, only the 
IH method will be used in this paper.

We now want to determine whether there are any ODEs of $\mathcal{S}_{re}$ 
that cannot be factorized by using the IH method.  Our main results are presented 
in the following proposition.\\

\noindent
{\bf Proposition 4.}  {\it Not all ODEs of $\mathcal{S}_{re}$ can be factorized 
by using the Infeld and Hull factorization method [34].}\\

\noindent
{\bf Proof}.  The semigroup $\mathcal{S}_{re}$ contains the ODEs of the 
form $\hat D y(x) = [ d^2 / dx^2 + B(x) d / dx + C(x) ] y (x) = 0$.  However, 
in general the functions $B(x)$, $C (x)$ and $y(x)$ may depend on a parameter 
$m$ that can be either integer or real, thus, we may write the above ODEs as 
$\hat D_m y(x, m) = [ d^2 / dx^2 + B_m(x) d / dx + C_m (x) ] y (x, m) = 0$.  
Let $\lambda$ be an eigenvalue of the operator $\hat D_m$, so that the 
corresponding eigenvalue equation becomes $\hat D_m y(x,m) = \lambda 
y (x, m)$, or in its explicit form is
\begin{equation}
y'' (x, m) + B_m (x) y' (x, m) + C_m (x) y(x, m) + \lambda y (x, m) = 0\ .
\label{S4eq5}
\end{equation}

Following the IH factorization method [34], Eq. (\ref{S4eq5}) is converted into 
a Sturm-Liouville equation and then cast into its {\it first canonical form}, which 
can be written as
\begin{equation}
{{d} \over {dx}} \left [ E_{s,m} (x) y'(x, m) \right ] + C_m (x) E_{s,m} (x) y (x, m) 
+ \lambda E_{s,m} (x) y (x,m) = 0\ ,
\label{S4eq6}
\end{equation}
where the Sturm-Liouville functions $p_m (x) = w_m (x) = E_{s,m} (x) = \exp{
[\int^x B_m (\tilde x) d \tilde x]}$ and $q_m (x) = C_m (x) E_{s,m} (x)$, with 
$\forall\ x \in (a,b) : p_m (x) = w_m (x) = E_{s,m} (x) > 0$, and with $a$ and 
$b$ being constants given by the boundary conditions.  

The procedure of converting the first canonical form into the second canonical 
form involves two transformations [34].  Then, the {\it second canonical form} 
of Eq. (\ref{S4eq6}) becomes  
\begin{equation}
\hat D_m^{\lambda} z (x,m) = z'' (x,m) + [\lambda + r (x, m)] z (x,m) = 0\ ,
\label{S4eq7}
\end{equation}
where $z(x,m)$ is the transformed dependent variable $y(x,m)$, $r(x,m)$ is 
given in terms of $p_m (x)$, $q_m (x)$ and $w_m (x)$ (see [37] for details).  
Expressing $p_m (x)$ and $w_m (x)$ by $E_{s,m} (x)$ and $q_m (x)$ in terms 
of $C_m (x)$ and $E_{s,m} (x)$, we get
\begin{equation}
r (x, m) = C_m (x) - {1 \over 2} \left [ B_m^{\prime} (x) + {1 \over 2} B_m^2 
(x) \right ]\ .
\label{S4eq8}
\end{equation}

Having obtained $r (x, m)$, it is shown that Eq. (\ref{S4eq7}) can be factorized 
if it can be replaced by each of the following equations [37]
\begin{equation}
\hat A_{+} (x, m+1) \hat A_{-} (x, m+1) z (\lambda, m) = [ \lambda - \chi 
(m+1) ] z (\lambda, m)\ ,
\label{S4eq9}
\end{equation}
and 
\begin{equation}
\hat A_{-} (x, m)  \hat A_{+} (x, m) z (\lambda, m) = [ \lambda - \chi (m) ] 
z (\lambda, m)\ ,
\label{S4eq10}
\end{equation}
where $\chi (m)$ is a function to be determined from $r (x, m)$ [37], and
\begin{equation}
\hat A_{\pm} (x, m) = \pm {{d} \over {dx}} + k (x, m)\ ,
\label{S4eq11}
\end{equation}
are the raising, $\hat A_{+}$, and lowering, $\hat A_{-}$, operators that give
\begin{equation}
z (\lambda, m+1) = \hat A_{-} (x, m+1) z (\lambda, m)\ ,
\label{S4eq12}
\end{equation}
and 
\begin{equation}
z (\lambda, m-1) = \hat A_{+} (x, m-1) z (\lambda, m)\ ,
\label{s4eq13}
\end{equation}
if $z (\lambda, m)$ is a solution of Eq. (\ref{S4eq7}).

According to IH [34], there are six types of factorization denoted as 
$A$, $B$, $C$, $D$, $E$ and $F$.  For each type, the functions $r 
(x, m)$, $k (x, m)$ and $\chi (m)$ are given in Table 17 of [34], 
which shows that the $x$-dependence of the functions $r (x, m)$ 
and $k (x, m)$ is different for each type of factorization, nevertheless, 
it remains fixed within the same type.  Similarly for $\chi (m)$ whose 
$m$-dependence is also fixed.  However, all three functions depend 
on some parameters that may change their values from one ODE to 
another.  There are examples of ODEs in Table 17 of [34] that 
correspond to many known ODEs of $\mathcal{S}_{sf}$ and also 
other examples of ODEs that we would classify as elements of 
$\mathcal{S}_{re}$.    

Now, using Eq. (\ref{S4eq8}), the function $r (x,m)$ can be obtained 
for any ODEs of $\mathcal{S}_{re}$.  However, since $B_m (x)$ and 
$C_m (x)$ are arbitrary functions of $x$, the function $r (x, m)$ may 
have in principle any form, not necessary one of the forms of $r (x,m)$ 
established by IH for their factorization types [34]; note that $r(x,m)$ for
each factorization type has its $x$-dependence fixed.  This shows that 
there are some ODEs of $\mathcal{S}_{re}$ that cannot be factorized 
by the IH method.  This concludes the proof.\\
   
The detailed calculations of casting the ODEs in their first and second 
canonical forms are lengthly and involve transformations that lead to a 
complicated form of $r (x, m)$ [37].  We show in the following proposition 
how these calculations can be significantly simplified.\\

 \noindent
{\bf Proposition 5.}  {\it Let $\hat D_m y(x,m) = [ d^2 / dx^2 + B_m (x) 
d / dx + C_m (x) ] y (x, m) = 0$, and let $\lambda$ be an eigenvalue of 
the operator $\hat D$, so that the corresponding eigenvalue equation is 
\begin{equation}
y'' (x, m) + B_m (x) y' (x, m) + C_m (x) y(x, m) + \lambda y_m (x) = 0\ .
\label{S4eq14}
\end{equation}
The following integral transformation 
\begin{equation}
y(x,m) = z(x,m) \exp \left [ - {1 \over 2} \int^x B_m (\tilde x) d \tilde x \right ]\ , 
\label{S4eq15}
\end{equation}
converts Eq. (\ref{S4eq14}) directly into its second canonical form $\hat 
D_m^{\lambda} z (x,m) = 0$ (see Eq. \ref{S4eq7}).}\\

\noindent
{\bf Proof}.  Applying the transformation given by Eq. (\ref{S4eq15}) to Eq. 
(\ref{S4eq14}), we obtain
\begin{equation}
z'' (x,m) + [ \lambda + r (x, m) ] z (x,m) = 0\ ,
\label{S4eq16}
\end{equation}
which is the second canonical form of Eq. (\ref{S4eq14}),
with
\begin{equation}
r (x, m) = C_m (x) - {1 \over 2} \left [ B_m^{\prime} (x) + {1 \over 2} B_m^2 
(x) \right ]\ .
\label{S4eq17}
\end{equation}
Note that the obtained result neither requires casting Eq. (\ref{S4eq14}) into 
its first canonical form nor using the original IH transformations [34].  This 
concludes the proof.\\
 
The results of Proposition 5 can be used to formulate the following corollaries.\\

\noindent
{\bf Corollary 6.} {\it All ODEs of $\mathcal{S}$ can be cast into their second 
canonical form.}\\

\noindent
{\bf Corollary 7.} {\it If the ODEs of $\mathcal{S}$ are cast into their second 
canonical form, then all their null Lagrangians are zero.}\\

It must also be noted that the null Lagrangians are zero for some ODEs of 
$\mathcal{S}_{sf}$, such as Airy, Hill, Mathieu, Weber, Wittaker and other 
similar ODEs. 

Let us apply the results of Proposition 5 to the Bessel equation by taking 
$m = \mu$, $B_{\mu} (x) = 1/x$, $C_{\mu} (x) = - \mu^2 / x^2$, $\lambda = 1$, 
and either $y (x,m) = J_{\mu} (x)$ or $y (x,m) = J_{-\mu} (x)$ as the linearly 
independent solutions.  By performing the transformation given by Eq. (\ref{S4eq15}),
the Bessel equation is cast in its second canonical form, and we find $r (x,m) = - 
(m^2 - 1/4) / x^2$, which gives $\hat A_{\pm} (x, \mu) = [ \pm d/dx - (\mu - 
1/2)/x]$, in agreement with the results given in Section 3.1.

We also want to point out that the IH factorization of the considered ODEs and 
the Lie group approach are closely related as shown by the following corollary.\\

\noindent
{\bf Corollary 8.} {\it The Lie group approach can only be established for those 
ODEs that can be factorized using the IH factorization method [34].  The results 
of Proposition 4 show that there are some ODEs of $\mathcal{S}_{re}$ that 
cannot be factorized, which means that the Lie group approach cannot be 
established for these ODEs.}\\

Our description of the Lie group approach showed that only ODEs of $\mathcal{S}_{sf}$ 
can be factorized and therefore the Lie group approach can be established for them.  
We also established that all ODEs of $\mathcal{S}_{re}$ can be converted into their 
second canonical forms, and that the general form of $r (x,m)$ for these ODEs can 
be obtained.  Despite the fact that $r (x,m)$ can be derived, our results showed that 
there are some ODEs of $\mathcal{S}_{re}$ that cannot be factorized using the IH 
method [34].

\section{Lagrange formalism versus Lie group approach}

In this paper, we have established the Lagrangian formalism based on standard 
and non-standard Lagrangians for all ODEs of $\mathcal{S}$, and demonstrated 
that the previously formulated Lie group approach is applicable to only some ODEs 
of $\mathcal{S}$.  Let $\mathcal{S}_{LL}$ be a sub-semigroup of all ODEs for 
which both the Lagrange formalism and the Lie group approach are established, 
and let $\mathcal{S}_{L}$ be another sub-semigroup of all remaining ODEs for 
which only the Lagrange formalism is established, with $\mathcal{S} 
= \mathcal{S}_{LL} \cup \mathcal{S}_{L}$.  The ODEs of $\mathcal{S}_{sf}$ 
seem to be special as they all belong to $\mathcal{S}_{LL}$, so it would be 
interesting to find deeper connections between the Lagrange formalism and 
the Lie group approach for these equations.  Similarly, for other ODEs of 
$\mathcal{S}_{re}$, which are also elements of $\mathcal{S}_{LL}$.  
Now, for the ODEs of $\mathcal{S}_{L}$ the underlying Lie groups are 
not yet known, so it would be interesting to find them.   

Possible solutions to the above problems involve developing the Lagrange 
formalism jointly with the Lie group approach, which would require 
replacing the configuration manifold $M$ by a manifold $G$ associated
with a given Lie group, so that $L: TG \rightarrow \mathcal{R}$.  The 
approach guarantees that the resulting Lagrangian $L$ is $G$-invariant, 
however, it may require deriving 'new' variational principles and 'new' 
Euler-Lagrange equations [38,39].  The resulting invariance of $L$ is its 
important property that is strongly related to the Lie group $G$.  Actually, 
the problem may be reversed and for every known Lagrangian its invariance 
with respect to rotations, translations and boots may indicate the presence of 
the underlying Lie groups [40,41].  In other words, the groups may be identified 
by investigating the Lagrangian invariance.  Moreover, the invariance of $L$ 
guarantees that the original equation derived from this Lagrangian preserved 
the same invariance [42].  It may be also interesting to explore relevance of 
the recently discovered particle-like structure of Lie algebra [43] to both the 
Lagrange formalism and the Lie group approach.

Let us now point out that the null Lagrangians that were identified in this 
paper as gauge transformations allow finding the gauge functions, and 
also likely gauge groups underlying these transformations.  The latter is 
not included in our paper as it will be a subject of future explorations.
Specifically, it would be of great interest to discovered relationships 
between the gauge groups and the Lie groups, whose irreps may be 
used to obtain the ODEs of mathematical physics with the special 
function solutions.

The above topics are of great importance in mathematical physics, and their 
exploration may give new connections between the Lagrange formalism and 
the gauge groups, and the Lie groups for the ODEs of $\mathcal{S}_{LL}$.  
It may also help finding Lie groups that underlie the ODEs of $\mathcal{S}_{L}$.  
However, these topics are out of the scope of the present paper but they 
will be investigated in future papers. 

\section{Conclusions}

A set of general second-order linear ordinary differential equations with non-constant 
coefficients was considered, and an algebraic structure (binary addition) was added 
to this set to form a semigroup.  An interesting result is that the operation of the 
semigroup can be used to produce many new differential equations, which are still
elements of the semigroup.  Among a few presented examples, novel Bessel equations
with their zero eigenvalues were obtained and discussed.  

The Lagrangian formalism was established for all ODEs of the semigroup.  We solved 
the inverse variational problem for these equations and derived standard and nonstandard 
Lagrangians.  Among the derived standard Lagrangians, we obtained the minimal, middle 
and maximal Lagrangians, and demonstrated that they are equivalent Lagrangians.  
Moreover, we showed that the minimal Lagrangian is the simplest and also the most 
fundamental, and that the other Lagrangians differ by the so-called null Lagrangians 
from the minimal one.  In our analysis of the null Lagrangians, we found out that these 
Lagrangians correspond to addition of the total derivative of a scalar function, which is 
a well-known result in calculus of variations [3,15], and in this paper was identified as
gauge transformations.  We also used the null Lagrangians to determine the gauge 
functions for all considered ODEs.   As long as non-standard Lagrangians are concerned, 
our results showed that these Lagrangians require that calculus of variations is amended 
by auxiliary conditions that are different for different ODEs.  This is a novel phenomenon 
in calculus of variations that is responsible for violation of the third Helmholtz conditions
for these Lagrangians.     

A Lie group approach was briefly described and compared to the Lagrange 
formalism established in this paper.  The comparison shows that the 
Lagrange formalism is more robust as it can be applied to all considered
equations, and that the Lie group approach is limited to the equations whose
solutions are the special functions of mathematical physics and some other 
similar equations.  We identified a sub-semigroup of ODEs that cannot be 
obtained by the Lie group approach.  Moreover, we also suggested alternative 
approaches to establish deeper connections between the Lagrangian formalism 
based on the standard and non-standard Lagrangians, the gauge (still to be 
determined) and the Lie group approach.

\bigskip\noindent
{\bf Acknowledgments}
We thank two anonimous referees for their valuable comments and 
suggestions that allows us to significantly improve the original version
of this paper.  This work was supported by the Alexander von Humboldt 
Foundation (Z.E.M.) and by the LSAMP Program at the University of Texas 
at Arlington (M.R.). 
%

\end{document}